\documentclass[twocolumn,showpacs,showkeys,prl,aps]{revtex4}
\usepackage{graphicx}

\newcommand{ \be }{\begin{equation}}
\newcommand{ \ee }{\end{equation}}
\newcommand{ \bea }{\begin{eqnarray}}
\newcommand{ \eea }{\end{eqnarray}}

\begin{document}

\title{Measuring Shear Viscosity Using Transverse Momentum Correlations
in Relativistic Nuclear Collisions}

\author{Sean Gavin$^a$ and Mohamed Abdel-Aziz$^b$}
\affiliation{
a) Department of Physics and Astronomy, Wayne State University, 666
W Hancock, Detroit, MI, 48202\\
b) Institut f\"ur Theoretische Physik, J.W. Goethe Universit\"at,
60438 Frankfurt am Main, Germany}
\date{\today}

\begin{abstract}
Elliptic flow measurements at RHIC suggest that quark gluon plasma
flows with very little viscosity compared to weak coupling
expectations, challenging theorists to explain why this fluid is so
nearly ``perfect''. It is therefore vital to find quantitative
experimental information on the viscosity of the plasma. We propose
that measurements of transverse momentum fluctuations can be used to
determine the shear viscosity. We use current data to estimate the
viscosity-to-entropy ratio in the range from 0.08 to 0.3, and
discuss how future measurements can reduce this uncertainty.
\end{abstract}


\maketitle
Measurements of elliptic and radial flow at RHIC are described by
viscosity-free hydrodynamics, indicating that the quark-gluon system
produced in these collisions is a nearly perfect liquid
\cite{kolb,derek0,HiranoGyulassy,Csernai:2006zz}. In particular, the
strong suppression of flow due to shear viscosity predicted by
weak-coupling transport calculations is not observed \cite{derek0}.
This result is exciting because a small viscosity relative to the entropy 
density of the system may indicate that
the system is more strongly coupled than expected: The collisional
shear viscosity is proportional to the mean free path, which is
shorter when the coupling is stronger.
But is the viscosity really small? Hirano {\it et al.}~point out
that color glass condensate formation may produce more elliptic flow
than considered in refs.~\cite{kolb,derek0}, requiring a larger
viscosity for agreement with data \cite{Hirano}.

We seek an experimental probe of viscosity that is independent of
elliptic flow. To that end, we propose that transverse momentum
correlation measurements can be used to extract information on the
kinematic viscosity,
\begin{equation}
\nu = \eta/Ts,
\label{eq:diffusion1}
\end{equation}
where $\eta$ is the shear viscosity, $s$ is the entropy density and
$T$ is the temperature. This ratio characterizes the strength of the
viscous force relative to the fluid's inertia and, consequently,
determines the effect of $\eta$ on the flow \cite{HiranoGyulassy}.
We argue that viscous diffusion broadens the rapidity dependence of
transverse momentum correlations, and then show how these
correlations can be extracted from measurements of event-by-event
$p_t$ fluctuations.

A number of experiments have studied transverse momentum
fluctuations at SPS and RHIC \cite{pruneau,STAR}. Interestingly, the
STAR collaboration reports a $60\%$ increase of the relative
rapidity width for $p_t$ fluctuations when centrality is increased
\cite{Adams:2005aw}. While the STAR analysis differs from the one we
propose, model assumptions provide a tantalizing hint that the
viscosity is small.

Any experimental information on the kinematic viscosity of high
energy density matter is vital for understanding the strongly
interacting quark gluon plasma. Theorists had long anticipated a
large collisional viscosity based on weak coupling QCD
\cite{QCDvisc} and hadronic computations \cite{hadroVisc}, with
values of $\eta/s$ roughly of order unity for both phases near the
crossover temperature $\sim 170$~MeV. Supersymmetic Yang Mills
calculations give the significantly smaller ratio $\eta/s = 1/4\pi$
in the strong coupling limit \cite{sQCD}. Lattice QCD calculations
of the shear viscosity will eventually settle the question of the
size of the viscosity near equilibrium \cite{lattice}. However, the
effective viscosity in the nonequilibrium ion-collision system may
differ from these calculations. In particular, plasma-instability
contributions can also explain the small viscosities in nuclear
collisions \cite{steffen}.

We begin by formulating a simple model to illustrate how shear
viscosity attenuates correlations due to fluctuations of the radial
flow. Next, we show how transverse momentum fluctuations can be used
to measure these correlations. We then demonstrate the impact of
viscosity on the rapidity distribution of fluctuations. Finally, we
explore the implications of current fluctuation data.

Before wading into the quark-gluon liquid, it is useful to recall
how shear viscosity affects the flow of more common fluids. In a
classic example of shear flow, a liquid is trapped between two
parallel plates in the $xy$ plane, while one plate moves at constant
speed in the $x$ direction. The fluid is pulled along with the
plate, so that $v_x$ varies with the normal distance $z$. The
viscous contribution to the stress energy tensor is then
\begin{equation}
T_{zx}=-\eta
\partial v_x/\partial z;
\label{eq:stress}
\end{equation}
see ref.~\cite{LL} for a general treatment.

Central nuclear collisions produce a high energy density fluid that
flows outward with an average radial  velocity $v_r$. In the
hydrodynamic description of these collisions, we typically assume
that $v_r$ varies smoothly with spacetime $(t, \mathbf x)$
and is the same for all collisions of a fixed impact parameter. For
central collisions, $v_r$ is radially symmetric.  More
realistically, small deviations ${\mathbf u}({\mathbf x})$ of the
radial flow occur throughout the fluid, varying with each ion-collision 
event. Such deviations occur, e.g., because the number and location of 
nucleon-nucleon subcollisions varies in each event. 

Viscous friction arises as neighboring fluid elements flow past each
other. This friction reduces $\mathbf u$, driving the velocity
toward the local average $v_r$.  The final size of the velocity
increment $\mathbf u$ depends on the magnitude of the viscosity and
the lifetime of the fluid.

In order to illustrate how the damping of radial flow fluctuations
depends on the viscosity of the fluid, we introduce a velocity
increment in the radial direction $u$ that depends only on the
longitudinal coordinate $z$ and $t$. Our aim is to determine the
linear response of the fluid to this perturbation. For simplicity,
we take the  unperturbed flow as slowly varying, and work in a
co-moving frame where $v_r$ locally vanishes. As in
(\ref{eq:stress}), the flow of neighboring fluid elements at
different radial speeds $u(z)$ produces a shear stress
\begin{equation}
T_{zr}=-\eta \partial u/\partial z. \label{eq:shear}
\end{equation}
This stress changes the radial momentum current of the fluid, which
is generally $T_{0r} = \gamma^2(\epsilon + p)v_r$ for energy density
$\epsilon$, pressure $p$, and $\gamma = (1-v^2)^{-1/2}$ \cite{LL}. The
perturbation
$u$ results in the change $g_t(\mathbf{x}) = \delta T_{0r}\approx
(\epsilon + p) u$ in the co-moving frame. On the other hand, the
energy-momentum conservation law $\partial_\mu T^{\mu\nu} =0$
implies $\partial g_t/\partial t = -\partial T_{zr}/\partial z$.

We combine these results to obtain a diffusion equation for the
momentum current
\begin{equation}
\left({{\partial}\over{\partial t}} - \nu\nabla^2 \right)g_t = 0
\label{eq:diffusion}
\end{equation}
to linear order, where the kinematic viscosity is given by (\ref{eq:diffusion1}),
since $\epsilon + p\approx Ts$ for small net baryon density. Observe
that (\ref{eq:diffusion}) applies for any fluctuation $\mathbf{g}_t$
for which $\mathbf{\nabla}\cdot {\mathbf g}_t =0$ \cite{LL}; our
physically-motivated radial $g_t(z,t)$ is a specific instance of
such a flow. Such shear modes are related to sound waves
(compression modes) but diffuse rather than propagate. Note that the
scale over which sound is attenuated $\Gamma_s = (4\eta/3+\zeta)/Ts$
depends on both shear and bulk viscosity \cite{LL,pratt}.

Viscosity tends to reduce fluctuations by distributing the excess
momentum density $g_t$ over the collision volume. This effect
broadens the rapidity profile of fluctuations. We write
(\ref{eq:diffusion})  in terms of the spatial rapidity $y =
1/2~\ln(t+z)/(t-z)$ and proper time $\tau=(t^2-z^2)^{1/2}$ to find
$\partial g_t /\partial\tau= (\nu/\tau^2)\partial ^{2}g_t/\partial
y^2$. A similar equation is used to study net  charge diffusion in
ref.~\cite{MohamedSean}, and we can translate many of those results to
the present context. Defining $V \equiv \langle(y-\langle y
\rangle)^2 \rangle = \int y^2g_t dy/\int g_t dy$ for $\langle
y\rangle = 0$, we compute the rapidity broadening
\begin{equation} \Delta V= \frac{2 \nu}{\tau_o}\left(1-
\frac{\tau_o}{\tau}\right), \label{deltaV}
\end{equation}
where $\Delta V\equiv{ V}-{ V}(\tau_o)$ for $\tau_o$ the formation time.

We extend this discussion to address a more general ensemble of
fluctuations by considering the correlation function
\begin{equation}
r_g = \langle g_t(\mathbf{x}_1)g_t(\mathbf{x}_2)\rangle - \langle
g_t(\mathbf{x}_1)\rangle \langle g_t(\mathbf{x}_2)\rangle.
\label{CcorrF2}
\end{equation}
In local equilibrium, $r_g$ has the value $r_{g,\,\rm eq}$. The
spatial rapidity dependence of $\Delta r_g\equiv r_g - r_{g,\, eq}$
is broadened by momentum diffusion. If the rapidity width of the
one-body density follows (\ref{deltaV}), then the width of $\Delta
r_g$ in the relative rapidity $y_r = y_1 - y_2$ grows from an
initial value $\sigma_0$ following
\begin{equation}
\sigma^2 = \sigma_0^2 + 2\Delta V(\tau_f), \label{fun}
\end{equation}
where $\tau_f$ is the proper time at which freeze out occurs.
This equation is entirely plausible, since diffusion spreads the rapidity
of each particle in a given pair with a variance $\Delta V$.
We then take
\begin{equation}\label{eq:ryy}
\Delta r_g(y_r,y_a) \propto
e^{-y_r^2/2\sigma^2-y_a^2/2\Sigma^2},
\end{equation}
where (\ref{fun}) gives the width in relative rapidity and the width in
average rapidity $y_a = (y_1+y_2)/2$ is $\Sigma$. We assume $\Sigma
\gg \sigma$ \cite{MohamedSean}. Observe that (\ref{fun}) and (\ref{eq:ryy})
are exact for our diffusion model \cite{MohamedSean}.

\begin{figure}
\centerline{\includegraphics[width=3.2in]{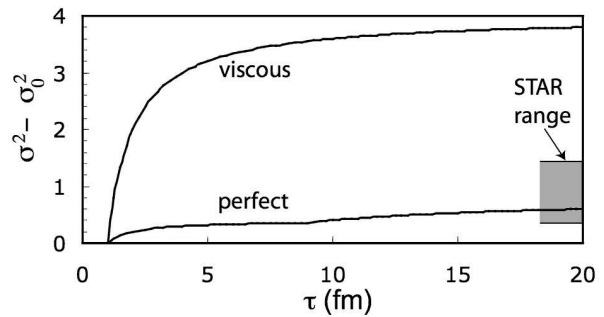}}
\caption[]{Rapidity spread vs. time for momentum diffusion computed
using (\ref{deltaV}) and (\ref{fun}) for the large viscosity
(viscous) and small viscosity (perfect) scenarios discussed in the
text. The gray area marks the range
extrapolated from data in ref.~\cite{Adams:2005aw} using (\ref{ptfluct*}).}
\label{fig:fig1}\end{figure}
Gyulassy and Hirano survey computations of the ratio of the shear
viscosity to the entropy and find that both the hadron gas and the
perturbative quark gluon plasma have $\eta/s\sim 1$, if one naively
extrapolates these calculations near $T_C$  \cite{HiranoGyulassy}.
These values correspond to $\nu =\eta/Ts$ roughly of order 1 fm for
$T_C=170$~MeV. On the other hand, they argue that the entropy
increase near $T_C$ reduces $\eta/s$ for a strongly interacting
plasma, perhaps to the supersymmetric Yang-Mills value $\eta/s =
1/4\pi$.

Motivated by these estimates, we show the increase of $\sigma$ given
by (\ref{deltaV}) and (\ref{fun}) as a function of proper time
$\tau$ for two extreme and highly schematic scenarios in
fig.~\ref{fig:fig1}. In the `perfect' scenario, we take $\nu \sim
0.1$~fm for the plasma and mixed phase, and $\nu\sim 1$~fm for the
hadronic phase. In the `viscous' scenario, we take $\nu \sim 1$~fm
for the entire evolution. In both cases, we assume that the
formation, hadronization, and freeze out times are 1~fm, 9~fm, and
20~fm respectively.

We stress that the rapidity width depends on the viscous diffusion
coefficient integrated over the collision lifetime. Comparing the
viscous and perfect scenarios in fig.~\ref{fig:fig1}, we see that
the largest contribution to this width comes from the earliest
times. Consequently, we expect measurements of this width to yield
information on the viscosity when the evolution is dominated by
partons.

Variation of the radial fluid velocity over the collision volume
induces correlations in the transverse momenta $p_t$ of particles
\cite{voloshin}. To describe such correlations, we divide the
inhomogeneous fluid into cells small enough to be uniform. Particles
emerging from cells of
different radial velocity $v_r$ are more likely to have different
$p_t$ than particles from the same cell. The number of
particles of momentum $\mathbf p$ in a cell at position $\mathbf x$
at the instant of freeze out is $dn = f(\mathbf{x},\mathbf{p})dpdx$,
where $dp\equiv d^3p/(2\pi)^{3}$ and $dx \equiv d^3x$. We take
$f(\mathbf{x},\mathbf{p})$ to be a Boltzmann
distribution corresponding to a fluid velocity ${\mathbf
v}(\mathbf{x})$ and a temperature $T(\mathbf{x})$ that vary with each
event.  A similar formulation is used in ref.~\cite{gavin}
to compute
nonequilibrium $p_t$ fluctuations. Here, we focus on central
collisions where local equilibrium is likely achieved.

To characterize the dynamic correlations of $p_t$, we use the
transverse momentum covariance
\begin{equation}\label{Cdef}
   {\cal C} =  \langle N\rangle^{-2}\langle \sum_{i\neq j} p_{ti}p_{tj}\rangle -\langle
   p_t\rangle^2,
\end{equation}
where  $i$ labels particles from each event and the brackets
represent the event average. The average transverse momentum is
$\langle p_t\rangle \equiv \langle \sum p_{ti}\rangle/\langle
N\rangle$.
This covariance vanishes in local equilibrium, where  the momenta
are uncorrelated and number fluctuations satisfy Poisson statistics.

This covariance is related to the spatial correlations of the
momentum current (\ref{CcorrF2}) by
\begin{equation}\label{CcorrF}
   {\cal C} = \langle N\rangle^{-2}\int  \Delta r_g(\mathbf{x_1}, \mathbf{x_2})
   dx_1dx_2.
\end{equation}
To obtain this result,
observe that $\langle N\rangle\langle p_t\rangle = \left\langle \int
p_{t} dn\right\rangle = \int \langle g_t(\mathbf{x})\rangle dx$,
where
\begin{equation}\label{momentum_density}
 g_t(\mathbf{x}) = \int f(\mathbf{x},\mathbf{p})p_t dp
\end{equation}
is the momentum current discussed earlier. Similarly, we write the
unrestricted sum $\sum p_{ti}p_{tj} = \int p_{t1}p_{t2} dn_1dn_2
=\int g_t(\mathbf{x}_1)g_t(\mathbf{x}_2) dx_1 dx_2$ and
average over events to find
\begin{eqnarray}\label{unres}
 \int  r_g dx_1dx_2
 &=&\langle \sum_{{\rm all}\, i,j} p_{ti}p_{tj}\rangle - \langle N\rangle^2\langle p_t\rangle^2
 \nonumber\\
    &=&\langle N\rangle^2{\cal C}+\langle\sum p_{ti}^2 \rangle;
\end{eqnarray}
the second equality follows from (\ref{Cdef}). In local equilibrium,
${\cal C}\equiv 0$ implies $\int r_{g,\,\rm eq}dx_1 dx_2   =
\langle\sum p_{ti}^2 \rangle$. Subtracting this term from
(\ref{unres}) gives (\ref{CcorrF}).

The correlation information probed by $\cal C$ differs from that
found in the multiplicity variance $R = (\langle
N^2\rangle-\langle N\rangle^2 - \langle N\rangle)/\langle
N\rangle^2$ \cite{PruneauGavinVoloshin}. As before, we write
$R = \langle N\rangle^{-2}\int  \Delta r_n dx_1dx_2$,
where $\Delta r_n = r_n-r_{n,\,\rm{eq}}$ and
\begin{equation}\label{NcorrF2}
r_n = \langle n(\mathbf{x}_1)n(\mathbf{x}_2)\rangle - \langle
n(\mathbf{x}_1)\rangle \langle n(\mathbf{x}_2)\rangle.
\end{equation}
The density correlation function (\ref{NcorrF2}) carries different
information than (\ref{CcorrF2}) because particle number is not conserved.
Density fluctuations evolve by the
full hydrodynamic equations, while $g_t$ follows diffusion.
The correlation function
probed by net charge fluctuations is discussed in ref.~\cite{MohamedSean}.

Viscosity information can be obtained from $\cal C$ as follows. The
broadening in rapidity of $\Delta r_g$ depends on the shear
viscosity via (\ref{fun}). Equation (\ref{CcorrF}) implies that the
rapidity dependence of $\Delta r_g$ can be measured by studying the
dependence of (\ref{Cdef}) on the rapidity window in which particles
are measured. We illustrate this acceptance dependence in fig.~2 for
our idealized scenarios by integrating (\ref{eq:ryy}) over the
interval $-\Delta/2 \le y_1,\, y_2 \le \Delta/2$; $\langle
N\rangle{\cal C}_{{}_\infty} $ is the value for large $\Delta$.
\begin{figure}
\centerline{\includegraphics[width=3.2in]{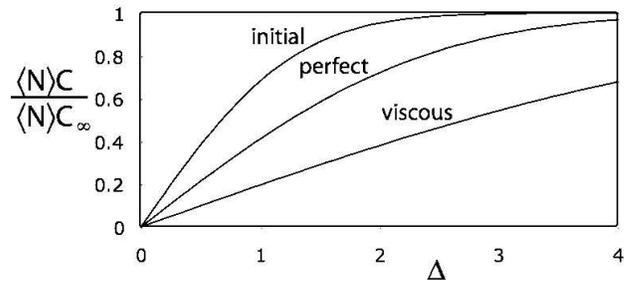}}
\caption[]{Rapidity dependence of the $p_t$ covariance (\ref{Cdef})
for the scenarios in fig.~1. The initial distribution has
$\sigma_0\sim 0.5$. }\label{fig:fig2}\end{figure}

The STAR analysis in ref.~\cite{Adams:2005aw} incorporates some of
these ideas and, intriguingly, finds a broadening in rapidity
together with a narrowing in azimuth for $p_t$ correlations in
central compared to peripheral collisions.
We will use the
rapidity information to estimate the viscosity. However, the
measured quantities differ sufficiently from $\cal C$ that this
estimate requires significant model assumptions. We therefore regard
the result only as a signal of our method's promise.

STAR employs the transverse momentum fluctuation observable
$\Delta\sigma_{p_t}^2$ to construct a correlation function as a
function of  rapidity and azimuthal angle. They find that near-side
correlations in azimuth are broadened in relative rapidity, with a
rapidity width $\sigma_*$ that increases from roughly 0.45 in the
most peripheral collisions to 0.75 in central ones
\cite{Adams:2005aw}. We estimate $\Delta\sigma_{p_t}^2/\langle
N\rangle \approx {\cal C} -\langle p_t\rangle^2R$. This quantity
therefore depends on both momentum current and density correlation
functions (\ref{CcorrF2}) and (\ref{NcorrF2}),
\begin{equation}\label{ptfluct*}
\Delta\sigma_{p_t}^2 \approx
   \langle N\rangle^{-1} \int  \{
   \Delta r_g
    -\langle p_t\rangle^2 \Delta r_n
    \}
    dy_1 dy_2.
\end{equation}
We can directly compare $\sigma_*$ to $\sigma$ in fig.~1 if $\Delta
r_g$ and $\Delta r_n$ have the same widths. Equation (\ref{fun})
then implies that the widths in central and peripheral collisions
satisfy $\sigma_c^2 - \sigma_p^2= 4\nu (\tau_{F,\, p}^{-1} -
\tau_{F,\,c}^{-1})$. Taking the freeze out times in a central
and peripheral collisions to be $\tau_{F,\,c}\sim$~20~fm and
$\tau_{F,\,p}\sim$~1~fm, we then find $\nu \sim 0.09$~fm. The value
$\tau_{F,\,p}\sim$~1~fm is reasonable, since  ref.~\cite{Adams:2005aw}
argues that the average participant path length is about 1~fm for
these peripheral collisions. We use (\ref{eq:diffusion1}) to find $\eta/s\sim 0.08$.

This result is remarkably close to the supersymmetric Yang Mills value
$1/4\pi$, and
is consistent with some hydrodynamic comparisons to elliptic flow
data \cite{derek0}.  However, we
must be cautious: If $\Delta r_g$ and $\Delta r_n$ have
different rapidity widths $\sigma$ and $\sigma_n$ then
their relation to $\sigma_*$ depends on the relative strength
of these contributions. Data in
ref.\ \cite{StarMult} may indicate that $\sigma_n$ is roughly twice $\sigma_*$.
Generally, $\sigma$ is bounded by $\sigma_n$
and $\sigma_*$, since  (\ref{ptfluct*}) implies
$\sigma_*^2 \approx \sigma^2 + \beta(\sigma^2-\sigma_n^2)$.
Although $\beta$ is not measured, the largest $\sigma$ may be is
$\sigma_n\sim 2\sigma_*$.  For the maximum value $\sigma = 2\sigma_*$, 
our dynamic assumptions
yield $\eta/s = 0.3$. Together, our estimates constitute an
uncertainty range for the viscosity-to-entropy ratio, $0.08 <
\eta/s<0.3$. 

In fig.~\ref{fig:fig1} we indicate the range of $\sigma_{*,\,
c}^2-\sigma_{*,\, p}^2$ for the most
central and peripheral STAR values  from ref.~\cite{Adams:2005aw}.
The gray band follows from the uncertainty 
in relating $\sigma$ to $\sigma_*$, i.e.,  $\sigma_* < \sigma
<2\sigma_*$, which greatly exceeds the experimental uncertainty.
Our `perfect' liquid curve falls above the bottom of this range because
we assume a large viscosity following hadronization. 

In summary, we find that shear viscosity can broaden the rapidity
correlations of the momentum current. This broadening can be
observed by measuring the transverse momentum covariance
(\ref{Cdef}) as a function of rapidity acceptance. Our rough
estimate from current data, $\eta/s\sim 0.08 - 0.3$, is small
compared to perturbative computations \cite{HiranoGyulassy}.
To reduce the uncertainty range, we suggest comparing
$\cal C$ to allow
more direct access to the momentum density correlation
function.  In principle, freeze out times
can be inferred from other measurements \cite{Lisa}.
Note that minijets, color glass, and other contributions to the particle
production mechanism influence the initial fluctuation spectrum and,
correspondingly, modify $\sigma_0$ in (\ref{fun}).
We assume that this contribution cancels in studying the
centrality dependence at a fixed beam energy.

The viscosity of a common fluid can be measured by applying a known
pressure and observing the resulting flow in a fixed geometry, e.g.,
a pipe. Alternatively, one can study the attenuation of high
frequency sound waves from a calibrated source. Efforts to compare
radial and elliptic flow measurements to viscous hydrodynamic
calculations are analogous to the first method \cite{derek0}. Our
observable $\mathcal C$ is in the spirit of ultrasonic attenuation.
The early dynamics produces a spectrum of fluctuations analogous to
sound waves that are attenuated by viscosity. We suggest that
experimenters  pursue both approaches to extract quantitative
viscosity information from ion collisions, since the geometry,
initial conditions, and probe parameters are all unknown.

We thank J.\ Dunlop for kindly bringing ref.~\cite{Adams:2005aw} 
to our attention, and R.\ Bellwied, M.\ Gyulassy, G.\ Moschelli, and C.\ Pruneau for
useful discussions. This work was supported in part by a U.S.
National Science Foundation PECASE/CAREER award under grant
PHY-0348559 (S.G.) and BMBF, GSI and DAAD (M.A-A.).

\vfill\eject
\end{document}